\documentclass[pra,twocolumn,showpacs,groupedaddress,nofootinbib]{revtex4}
\usepackage{amsmath,amssymb,amsfonts}
\usepackage{graphicx}
\usepackage{subfigure}
\usepackage{txfonts}
\usepackage{color}
\usepackage[capitalize]{cleveref}

\begin{document}

\title{Resonant transfer of large momenta from finite duration pulse sequences}

\author{J.~Fekete$^1$, S.~Chai$^1$, S.~A.~Gardiner$^2$, and M.~F.~Andersen$^1$}
	\email{mikkel.andersen@otago.ac.nz}
	\homepage{http://www.physics.otago.ac.nz/nx/mikkel/home-page.html}
\affiliation{\small$^1$ {The Dodd-Walls Centre for Photonic and Quantum Technologies, Department of Physics, University of Otago, Dunedin, New Zealand }\\
\small$^2$ Joint Quantum Centre (JQC) Durham-Newcastle, Department of Physics, Durham University, Durham DH1 3LE, United Kingdom}
\pacs{}

\begin{abstract}
We experimentally investigate the atom optics kicked particle at quantum resonance using finite duration kicks. Even though the underlying process is quantum interference it can be well described by an $\epsilon$-pseudoclassical model. The $\epsilon$-pseudoclassical model agrees well with our experiments for a wide range of parameters. We investigate the parameters yielding maximal momentum transfer to the atoms and find that this occurs in the regime where neither the short pulse approximation nor the Bragg condition is valid. Nonetheless, the momentum transferred to the atoms can be predicted using a simple scaling law, which provides a powerful tool for choosing optimal experimental parameters. We demonstrate this in a measurement of the Talbot time (from which $h/M$ can be deduced), in which we coherently split atomic wave-functions into superpositions of momentum states that differ by 200 photon recoils. Our work may provide a convenient way to implement large momentum difference beam splitters in atom interferometers.  
\end{abstract}

\maketitle

\section{Introduction and motivation}
The atom optics $\delta$-kicked particle is a paradigmatic system for experimental studies of quantum chaos and classical-quantum correspondence \cite{Oskay2000,Summy2001,Wu2009,Hoogerland2012,Summy2016}. It consists of laser cooled atoms exposed to a periodically pulsed standing wave (SW) laser field, tuned far off-resonant to relevant atomic transitions. A purely \textit{quantum} phenomenon in such systems is the appearance of quantum resonances (QR) which are a result of self-revivals of the atomic wave-function due to the matter-wave Talbot effect \cite{Phillips1999}. QRs lead to linear / ballistic growth in the root-mean-square momentum imparted to the atoms with the number of SW pulses \cite{Oskay2000,Sadgrove2005,Ryu2006}. The nonlinear dynamics of the $\delta$-kicked particle enables measurements with sub-Fourier precision \cite{Cubero} both in the vicinity \cite{Talukdar2010,Prentiss2009} and away from QR \cite{Szriftgiser2002}. 
In this context, it is very appealing to realize the large momentum transfer (LMT) of QR as a ''beam splitter'' (BS) in atom interferometry, as the sensitivity of atom interferometers grows with the momentum difference between the arms. This would allow for applications in high precision metrology such as measurements of $h/M$ \cite{Cadoret2008} etc. 
A number of atom interferometers today use series of low order Bragg diffraction pulses to realize LMT BS \cite{Kasevich2011,Tino2015}. Using QR bears similarities to this approach since it achieves LMT through consecutive low order diffractions. Compared to a single short pulse BS \cite{Phillips1999,Sleator2009} consecutive pulses can yield enhanced momentum transfer to the atoms. Interestingly, the pulse durations we consider are lower by typically two orders of magnitude compared to Bragg pulses \cite{Kasevich2011,Altin2013,Tino2015}. 
Using QR thereby reduces the interaction with the SW light, which is a potential source of systematic errors, noise, and decoherence in atom interferometers. Thus, QR is a promising approach for implementing LMT beam splitting processes in an interferometer. 

The $\delta$-kicked particle description is valid when the motion of atoms can be neglected during the SW pulses (Raman-Nath approximation). The finite pulse duration often needs to be accounted for numerically \cite{Oskay2000,Sadgrove2005} when comparing experiments to theoretical predictions. Furthermore, for a given SW power the maximal momentum transfer can be achieved when the SW pulse duration violates the Raman-Nath condition \cite{Sleator2009,Daszuta2012}. This has motivated the recent development of an $\epsilon$-pseudoclassical model which accounts for the finite pulse duration effects during QR \cite{Gardiner2016}. 
Here, we provide the first experimental test of the $\epsilon$-pseudoclassical model which is capable of predicting the momentum transfer to a group of atoms from finite duration SW pulses. We find that the model agrees well with our experiments for a surprisingly large range of pulse durations. For relevant parameters the width of the momentum distribution can be predicted using a simple scaling law. This is a powerful tool that allows for easy optimization of experimental parameters. We demonstrate this by a measurement of the Talbot time in which we split atoms into coherent superpositions of momentum states that differ by up to 200 photon recoils. 
For the regime where our LMT BS is realized, neither the Raman-Nath approximation nor the Bragg condition holds.

\begin{figure}[t!]
	\includegraphics[width=\linewidth]{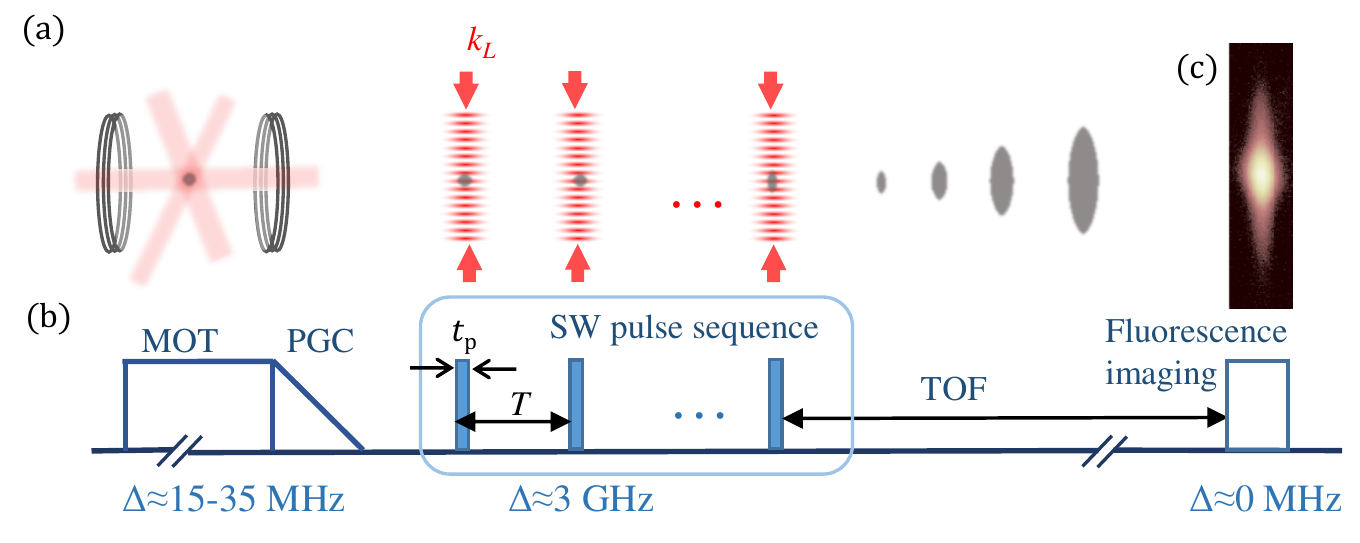}
	\caption{(Color online)
		Scheme of the experimental sequence. (a) Laser trapping and cooling; SW pulse sequence; expansion of the atomic cloud during time-of-flight; fluorescence imaging, (b) intensity and detuning ($\Delta$) of light, during the same sequence, not to scale, (c) fluorescence image. 
		\label{fig:scheme}}
\end{figure}

\section{Experimental sequence}
Our experimental sequence is depicted schematically in Fig.~\ref{fig:scheme}. We trap a cloud of $ ^{85}$Rb atoms in a magneto-optical trap (MOT): subsequent polarization gradient cooling (PGC) leaves the atoms at $\sim6.4$~$\mu$K in the $|5^2S_{1/2}, F=2\rangle$ state. We then apply the SW pulse sequence. The SW field is a laser beam retro-reflected by a mirror in the horizontal plane, $\sim40$~MHz red detuned from the $|5^2S_{1/2}, F=3\rangle \rightarrow |5^2P_{3/2}, F=4\rangle$ transition. For the initial internal state this light is off-resonant with $\sim3$~GHz red detuning. We apply $N$ SW pulses of duration $t_\mathrm{p}$ and period $T$. After the pulse sequence the atomic cloud freely expands for 9.9~ms time-of-flight (TOF), and finally we take a fluorescence image of the atomic distribution. 

\section{Theory}
To account for the finite pulse durations we use the $\epsilon$-pseudoclassical model described in \cite{Gardiner2016} (conceptually similar to the approach taken by Wimberger \textit{et al.}~\cite{Wimberger2004}). The model is as follows. We consider the 1D atomic motion along the SW axis. 
If the kicking period $T$ is an integer multiple ($L$) of the Talbot time $T_{T} = 4\pi M / \hbar K^{2}$ (quantum resonance), then the one period time evolution is governed by the Floquet operator: 
\begin{equation}\label{eq:Floquet}
\hat{F} = \exp
\left(
-\frac{i}{\hbar}
\frac{ \hat{p}^{2}}{2M}
[L T_{T}-t_\mathrm{p}]
\right)
\exp
\left(
-\frac{i}{\hbar}\left[
\frac{ \hat{p}^{2}}{2M}
-\frac{V_{\mathrm{d}}}{2}\cos(K\hat{x})\right]t_\mathrm{p}
\right).
\end{equation}
The right exponential term is the time evolution during the SW pulse, and the left the free evolution between pulses. $M$ is the atomic mass, and $K=2k_{L}$, with $k_{L}$ the SW laser wave number. $\hat{x}$ and $\hat{p}$ are position and momentum operators, respectively, and $V_{\mathrm{d}}$ is the SW potential depth. 

We rewrite Eq.~(1), taking advantage of two properties. Firstly, due to the spatial periodicity of the SW potential the quasimomentum \cite{beta} is conserved, so we restrict our analysis to manifolds of a given quasimomentum $\beta$ \cite{Bach2005}. Secondly, we use the revivals that a spatially periodic wave-function undergoes after free space evolution for duration $T_T$ \cite{Phillips1999}. Eq.~(1) can be rewritten in terms of rescaled dimensionless quantities $\epsilon = \hbar K^{2} t_\mathrm{p}/M$, $\hat{\theta}=K\hat{x}$, $\hat{\mathcal{J}} = \hat{p}\epsilon/\hbar K$ \cite{J}, and $\tilde{V} = V_{\mathrm{d}}t_\mathrm{p}\epsilon/2\hbar$ as \cite{Gardiner2016}:  \begin{equation}\label{eq:FloquetDimless}
\hat{F} = \exp
\left(
-\frac{i}{\epsilon}\left[
-\frac{\hat{\mathcal{J}}^{2}}{2} + \hat{\mathcal{J}}4\pi L \beta 
\right]
\right)
\exp
\left(
-\frac{i}{\epsilon}\left[
\frac{\hat{\mathcal{J}}^{2}}{2}
-\tilde{V}\cos(\hat{\theta})\right]
\right).
\end{equation} 
In this form of the Floquet operator new quantities appear at different positions. The role of $\hbar$ is played by $\epsilon$ which depends on $t_\mathrm{p}$, as also revealed by the commutation relation, $[\hat{\theta},\hat{\mathcal{J}}] = i\epsilon$. The apparent duration of both exponential operators is one dimensionless time unit. 
One often speaks of quantum dynamics converging to classical dynamics in the limit of $\hbar \rightarrow 0$. In the $\epsilon$-pseudoclassical model the dynamics of Eq.~(2) is  approximated with its classical counterpart assuming $\epsilon \ll 1$. 
\begin{figure}[t!]
	\includegraphics[width=\linewidth]{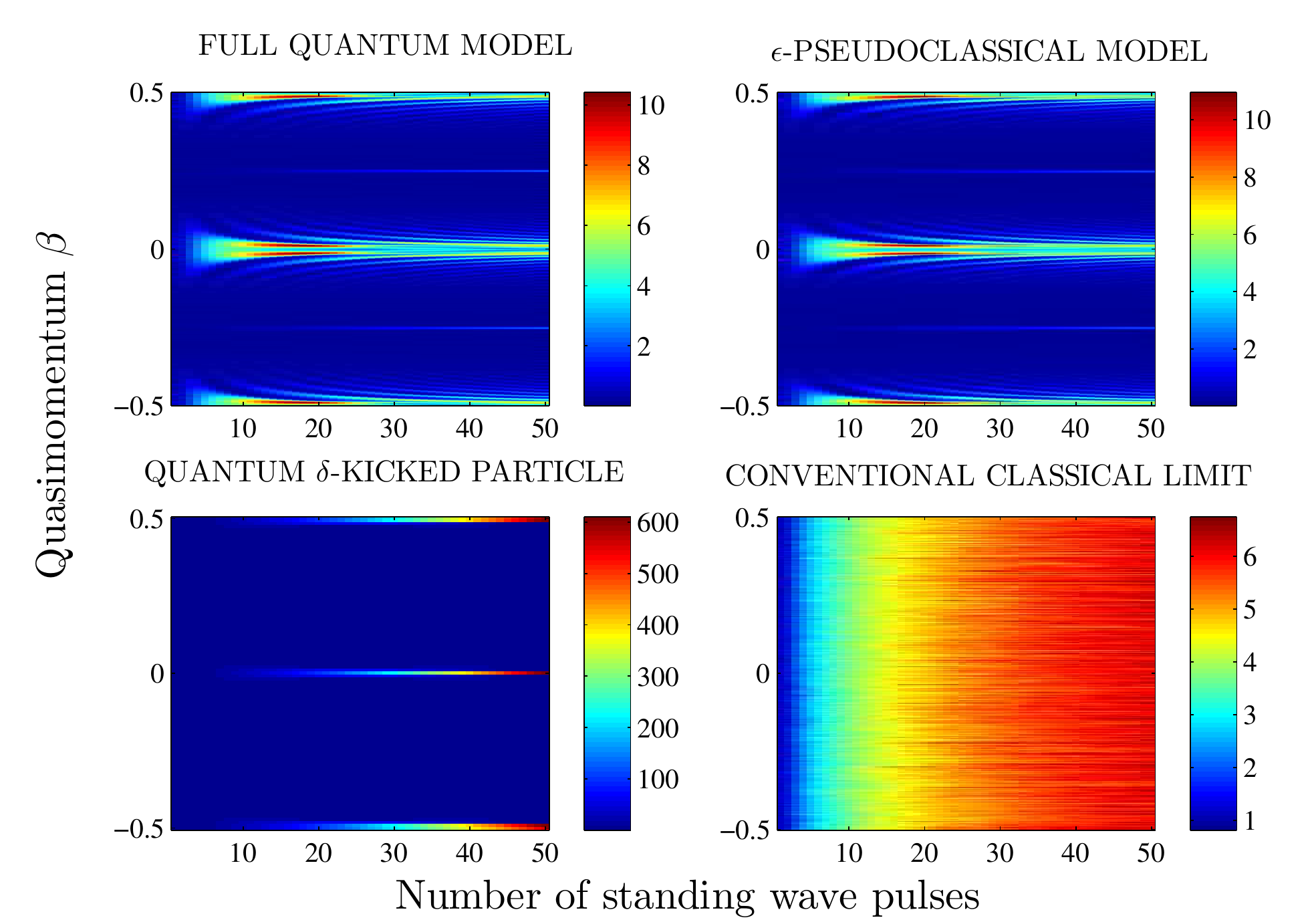}
	\caption{(Color online)
		Comparison of $\langle\mathcal{J}^2/2\rangle$ (shown in color coding, see color bars) in different models for $\epsilon=0.1, \tilde{V}=1$, as described by Eqs.~(1-3).
		\label{fig:compare_models}}
\end{figure} 
The effective classical dynamics is governed by the \emph{effective classical Hamiltonians} extracted from Eq.~(2). These are
$H_{1} = \mathcal{J}^{2}/2 - \tilde{V}\cos(\theta) $ and
$H_{2} = -\mathcal{J}^{2}/2 + \mathcal{J}4\pi L\beta$. $H_{1}$ still has the form of a pendulum, which is exactly solvable in terms of Jacobi elliptic functions. Solving Hamilton's equations of motion for $H_{2}$ yields the following map, which gives $\theta_2$ and $\mathcal{J}_2$ after the evolution under $H_{2}$ in terms of $\theta_1$ and $\mathcal{J}_1$ before it: 
\begin{subequations}
	\begin{align}
\theta_2  &= \theta_1 - \mathcal{J}_1 + 4\pi L \beta , \label{eq:map1}
\\ 
\mathcal{J}_2 &= \mathcal{J}_1 .\label{eq:map2}
\end{align}\label{eq:map}
\end{subequations} 
It is important to note that the $\epsilon$-pseudoclassical model is not the classical limit of our physical system. On the contrary, it consists of mapping the system onto a different classical system that captures the quantum dynamics of the actual system. This is illustrated in Fig.~\ref{fig:compare_models} where $\langle\mathcal{J}^2/2\rangle$ (which plays the role of the mean kinetic energy in the $\epsilon$-pseudoclassical model) is plotted as a function of pulse number and initial (quasi-) momentum, computed using different models. For details on the numerical methods see Appendix C. The $\epsilon$-pseudoclassical model is in quantitative agreement with the full quantum model (Eq.~(1)) for the parameters $\epsilon=0.1, \tilde{V}=1$ used. Neither the $\delta$-kicked particle model nor the classical model using the Hamiltonians corresponding to the classical limit of Eq.~(1) agrees with the full quantum model. 

Resonant transfer of kinetic energy to the atoms happens close to $\beta=0$ and to integer multiples of 1/2. It leads to quadratic increase in energy with the number of SW pulses up to a point ($N\approx5$ pulses in Fig.~\ref{fig:compare_models}) after which the energy transfer ceases. The strong dependence of QR on $\beta$ and the limit on the achievable kinetic energy indicates the challenges of transferring large momentum to a finite temperature gas. For instance efficient transfer of momentum to $>95\%$ of the atoms requires an initial momentum width below $0.2\hbar k_L$ for parameters of Fig.~\ref{fig:compare_models} and N = 7. This can be achieved using a Bose-Einstein condensate or by velocity selection \cite{Phillips1999,Tino2015}. 
For the quantum $\delta$-kicked particle the quadratic increase in energy is unlimited, however LMT is not feasible due to the increase in required laser power with $N$. 
$H_1$, $H_2$, and Eq.~(3) provide insight into the advantage of using consecutive finite duration pulses. For a single pulse the transferred kinetic energy is bounded by the SW potential depth. This can be directly seen from the pendulum Hamiltonian $H_1$: when the particle reaches the bottom of the potential it will start losing energy by climbing the next hill. Considering the $\beta=0$ subspace and Eq.~(3) we see that the evolution governed by $H_2$ does not change the scaled momentum $\mathcal{J}$ (and therefore not the actual momentum) but it changes the position in opposite direction to the momentum. This means that after the particle has rolled down a hill, picking up kinetic energy, the free space evolution by $H_2$ may bring it back up the hill, thereby allowing it to roll down the hill again during the next evolution under $H_1$, permitting it to pick up more energy and momentum. This way the particle can gain significant energy by rolling down the same hill many times. 
The origin of this apparent backwards motion is in the matter-wave Talbot effect. We note that the free space evolution in Eq.~(1) is for a duration $L T_T-t_\mathrm{p}$. Since a spatially periodic wave-function revives every $T_T$, evolving for a duration $T_T-t_\mathrm{p}$ is equivalent to a free space evolution of $t_\mathrm{p}$ backwards in time. In the $\epsilon$-pseudoclassical model this translates to the position changing in the opposite direction of the momentum. Note, that if $\beta \neq 0$, we get additional motion during the free flight ($4\pi L\beta$ term) suppressing the resonance effect. 

\begin{figure}[t!]
	\includegraphics[width=0.9\linewidth]{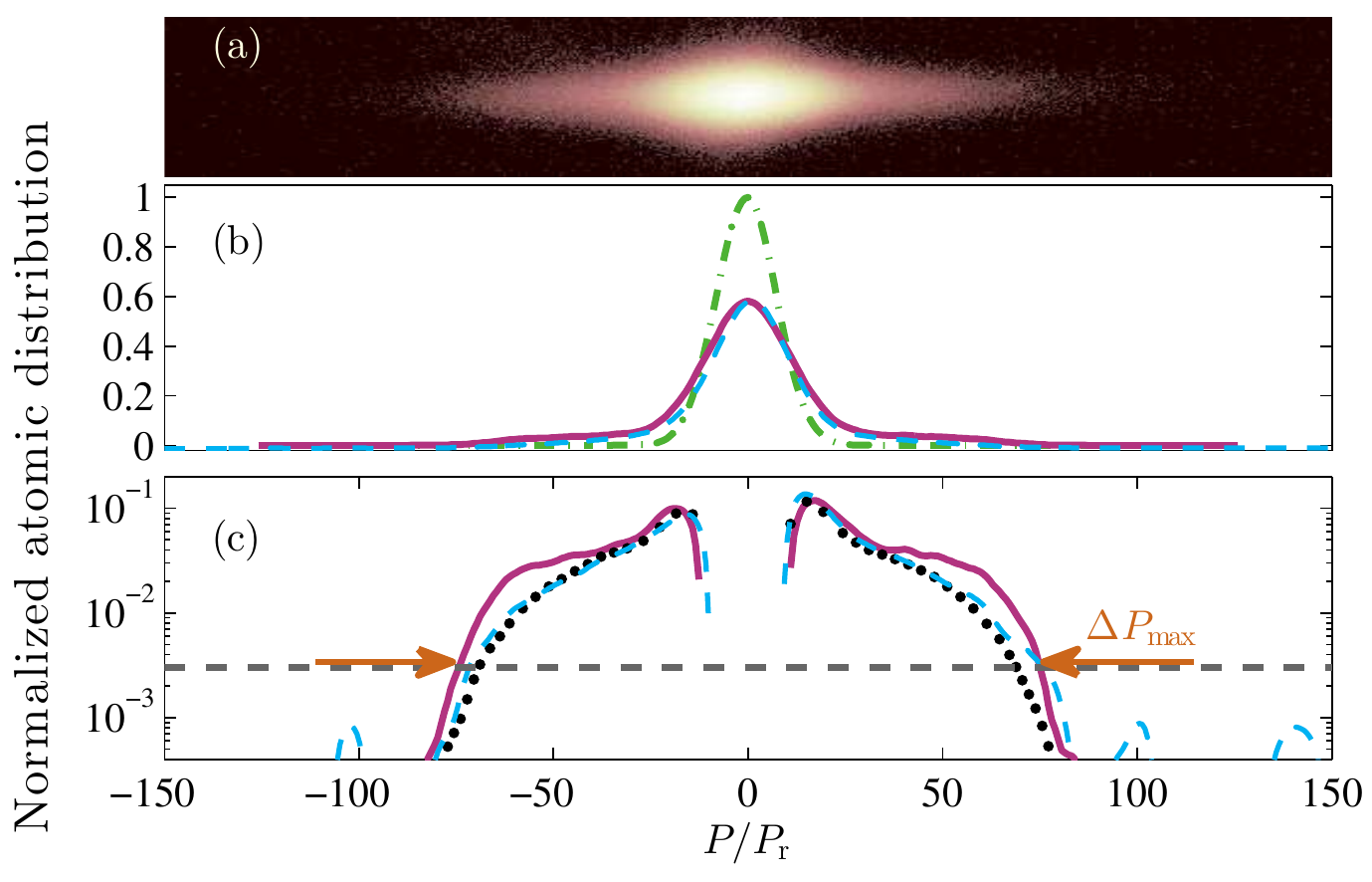}
	\vspace{-4mm}
	\caption{(Color online) Comparison of experiments and the $\epsilon$-pseudoclassical model. (a) Average of 10 fluorescence images with logarithmic color map used for calculating momentum distributions. (b) Measured and smoothed momentum distribution without (dash-dotted line) and with SW pulse sequence (dashed line), and $\epsilon$-pseudoclassical model for the same parameters (solid line). (c) Difference curves (SW $-$ no SW). Dashed line is measured data, solid line is from the model and dotted line is from the model including a range of potential depths seen by the atoms.}\label{fig:crossections}
	\vspace{0mm}
\end{figure} 

\section{Results}
\subsection{Validation of the $\epsilon$-pseudoclassical model}
To compare measurements with the $\epsilon$-pseudoclassical model, we investigate the cross-sectional atomic distributions along the SW beam axis obtained from averaging 10 repetitions of the experimental sequence (see Fig.~\ref{fig:crossections}). The cross-sectional distributions in the case of no SW light and of a sequence of $N=6$ pulses are plotted (dash-dotted and dashed lines, respectively) for parameters $t_\mathrm{p}=250$~ns, $V_{\mathrm{d}}/h=7.24$~MHz and $L=1$. The total time in the two cases was the same. We deduce the momentum, in units of photon recoil momentum ($P_{r}=\hbar k_L$), from the images using the time-of-flight. The atomic distributions are broadened due to the SW kicks and a fraction of atoms undergo LMT. To observe the distribution at the wings more carefully, we subtract the distribution with no SW pulses. This difference is shown in Fig.~\ref{fig:crossections} (c) in logarithmic scale after smoothing (dashed line), see Appendix D. 
We determine the maximum momentum difference of the atomic distribution, $\Delta P_\mathrm{max}$ at a universal threshold value, indicated with the horizontal dashed line in Fig.~\ref{fig:crossections} (c). The threshold value is chosen to be above the measurement noise level and it is a fixed value for all measurements. Solid lines in Fig.~\ref{fig:crossections} are calculations with the $\epsilon$-pseudoclassical model. For these calculations the initial width of the atomic distribution and $V_{\mathrm{d}}$ were chosen as best fit parameters, and they are within $25\%$ of the estimated value determined using measured quantities (see Appendix A and B for the experimental parameters).
 \begin{figure}[!t]
  	\includegraphics[width=\linewidth] {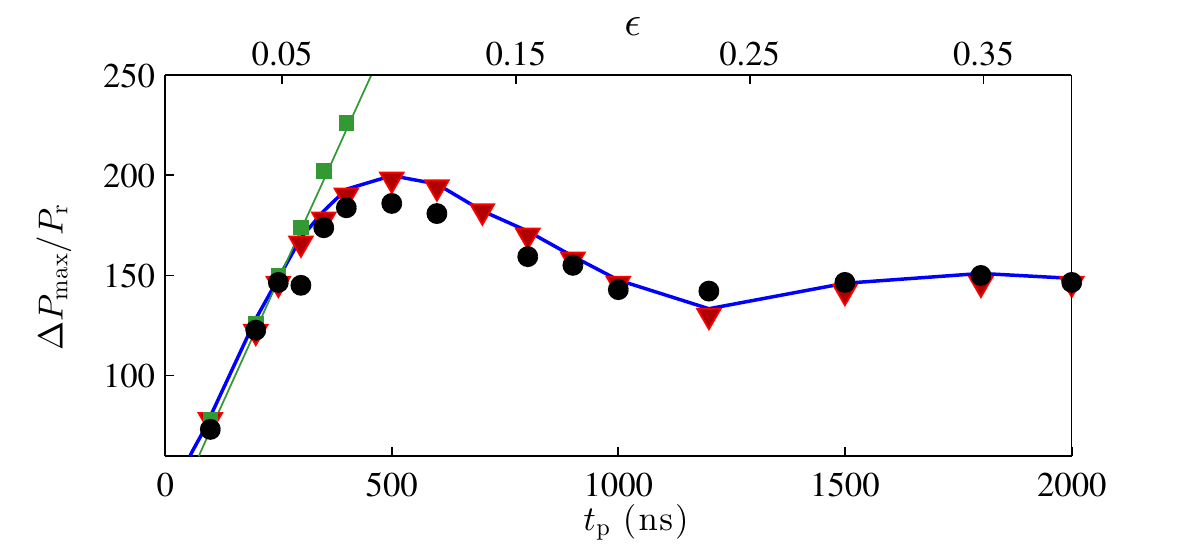}
  	\vspace{-6mm}
  	\caption{ (Color online)  $\Delta P_{\mathrm{max}}$ dependence on $t_\mathrm{p}$. 
  		Measured data (circles), quantum $\delta$-kicked particle (squares with linear fit), full quantum model (triangles), and $\epsilon$-pseudoclassical model (thick solid line) for the same parameters. 
  		\label{fig:Pmax_vs_tau}	}
  \end{figure} 
The $\epsilon$-pseudoclassical model proved to be a great tool to understand various effects that may come into play during the LMT process. Due to the flexibility of the Monte-Carlo simulations we could easily include various effects that imitate possible physical processes that atoms undergo during their interaction with the SW light sequence without a significant increase in the computational time. Such effects are phase fluctuations of the SW field or spontaneous photon scattering resulting in incoherent momentum exchange. We also modeled the effect of non-uniform potential depth over the atomic cloud, which we found to be the dominant effect for the small deviation observed in the upper part of the shoulders in Fig.~\ref{fig:crossections}~(c) (see Appendix B for details). 
Including these variations yielded only a small difference in $\Delta P_\mathrm{max}$, therefore we omit them in the following.
In Fig.~\ref{fig:Pmax_vs_tau} we compare $\Delta P_\mathrm{max}$ values calculated with the $\epsilon$-pseudoclassical model, the full quantum model, and the $\delta$-kicked particle model, to experimental data. We plot $\Delta P_\mathrm{max}$ values for a series of pulse durations $t_\mathrm{p}$, with $V_{\mathrm{d}}/h=7.24$~MHz and $N=6$. The experimental data (circles) are in good agreement with the $\epsilon$-pseudoclassical (thick solid line) and full quantum models (triangles). Here, the range up to $\epsilon\approx0.39$ ($t_\mathrm{p}= 2~\mu$s) is shown, but the agreement holds up to $\epsilon\approx1$. This is surprising, as $\epsilon \ll 1$ was assumed for the $\epsilon$-pseudoclassical model to be valid. 
In contrast, the $\delta$-kicked particle model (squares with linear fit) that predicts linear growth deviates significantly for $t_\mathrm{p}>250$~ns. 
We note that, when a combination of parameters is large (typically when both $t_\mathrm{p}>2~\mu$s, $V_{\mathrm{d}}/h>7$~MHz, and $N>12$), we observed significant discrepancies between the experimental data and the models. This could be due to phase instability of the SW. 
\begin{figure}[!t]
	\centering
	\includegraphics[width=\linewidth] {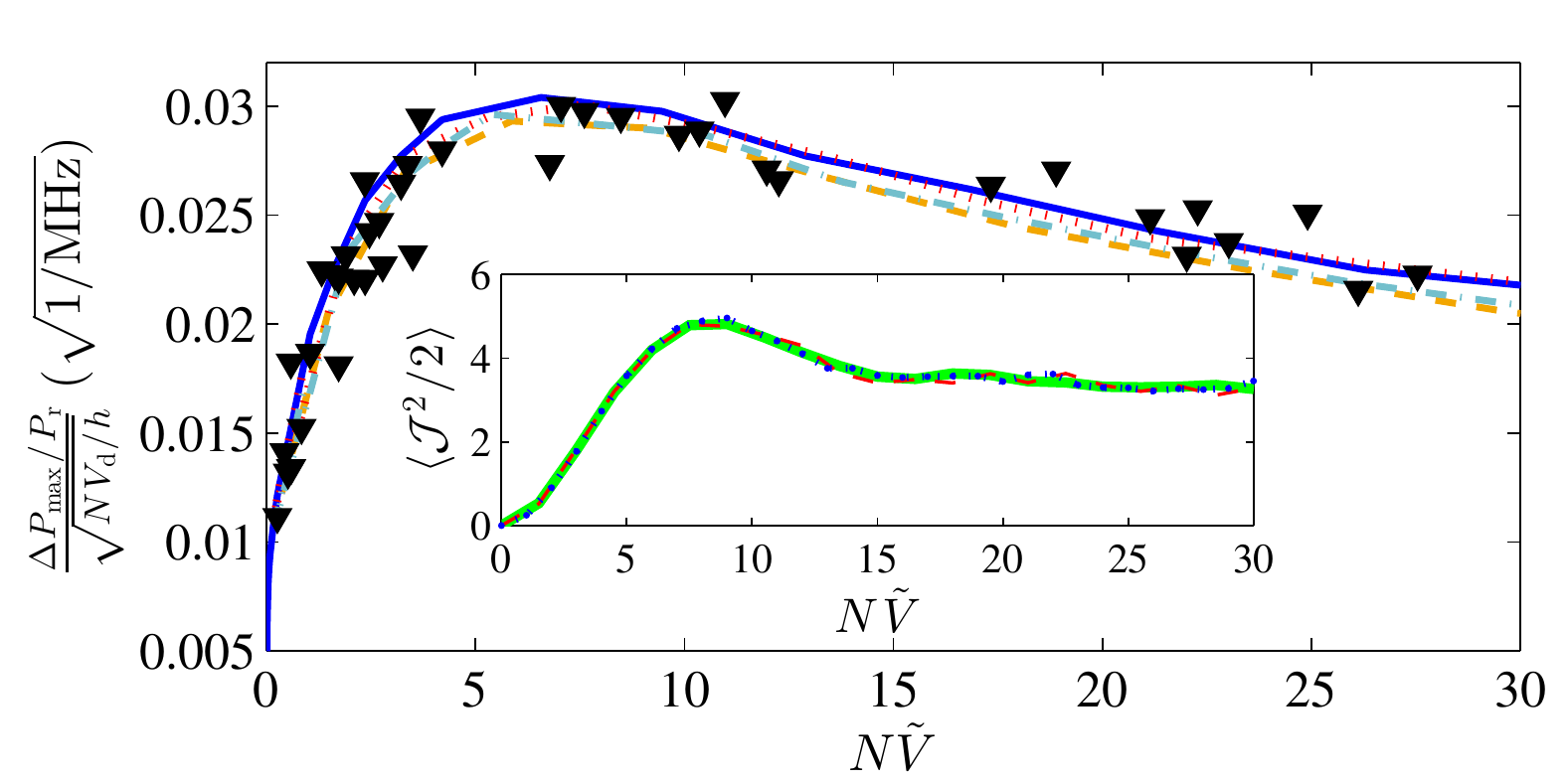}
	\vspace{-6mm}
	\caption{ (Color online)
		Scaling law. The lines show calculations from the $\epsilon$-pseudoclassical model with $t_\mathrm{p}$ scanned at 
		$N=6, V_{\mathrm{d}}/h=7.24$~MHz (solid line), 
		$N=9, V_{\mathrm{d}}/h=6.64$~MHz (dashed line), 
		$N=10, V_{\mathrm{d}}/h=3.47$~MHz (dotted line), 
		$N=14, V_{\mathrm{d}}/h=2.57$~MHz (dash-dotted line). Triangles show measured data within the range of  $N=5- 12$ and $V_{\mathrm{d}}/h=2.26-7.24$~MHz. 
		Inset shows scaling law for $\langle\mathcal{J}^2/2\rangle$ with $N=10, 50, 100$, and $V_{\mathrm{d}}/h=18.5, 3.7, 0.62$~MHz, respectively for $\beta=0$.
		\label{fig:scaling_law}}
\end{figure}

\subsection{Scaling law}
For the $\beta = 0$ subspace, \cite{Gardiner2016} found that $\langle\mathcal{J}^2/2\rangle$ followed a universal curve if the horizontal axis was scaled appropriately. The general form of this scaling law includes variations in $V_{\mathrm{d}}$ as shown in the inset of Fig.~\ref{fig:scaling_law}. We wish to verify this scaling law experimentally. Since we use a thermal gas, a measurement of the mean kinetic energy or $\langle\mathcal{J}^2/2\rangle$ would be skewed by the large proportion of atoms with $\beta$ away from resonances. However, for a wide range of parameters $\Delta P_{\mathrm{max}}$ is dominated by the resonant atoms, so it is intriguing to investigate if an equivalent scaling law exists for $\Delta P_{\mathrm{max}}$. Fig.~\ref{fig:scaling_law} shows an equivalent scaled graph for $\Delta P_{\mathrm{max}}$ assuming that $\langle\mathcal{J}^2/2\rangle$ of the $\beta =0$ subspace is proportional to $\Delta P_{\mathrm{max}}^2$. The scaling law is transformed (see Appendix E for details) to make the vertical axis independent of $t_\mathrm{p}$, such that we can use Fig.~\ref{fig:scaling_law} to determine the optimal value of $t_\mathrm{p}$. 
We find that, for the parameters chosen, $\epsilon$-pseudoclassical calculations and experimental data approximately follow a universal curve. If one chooses parameters such that $\Delta P_{\mathrm{max}}$ is not determined by the resonant atoms (e.~g.~when the characteristic shoulder in Fig.~\ref{fig:crossections} is below our threshold line), then we naturally see deviations from the universal curve.  

\subsection{Measurement of the Talbot time}
\begin{figure}[t!]
	\centering
			\includegraphics[width=\linewidth]{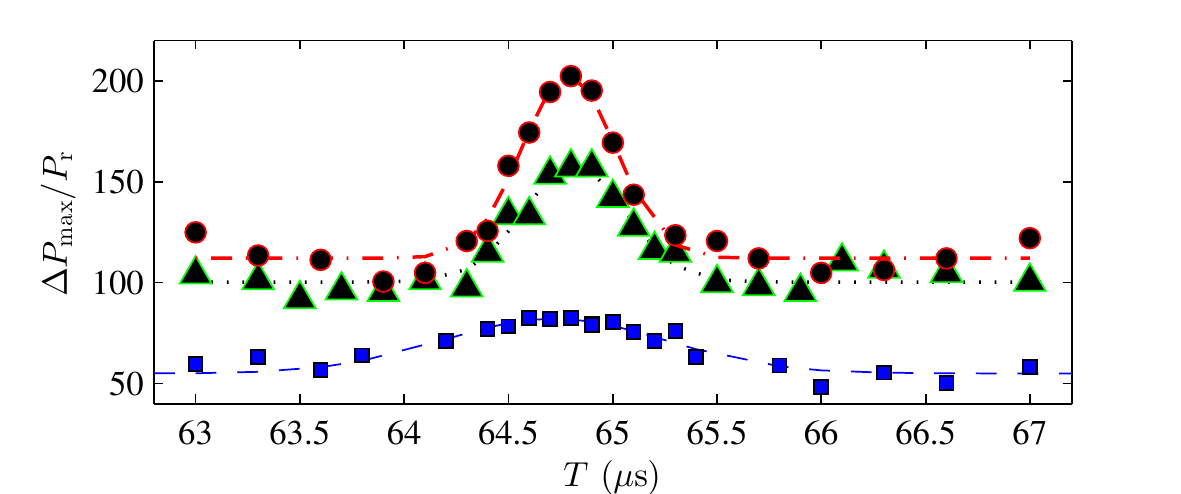}
	\caption{ (Color online)
		Measurements of the Talbot time. Optimum performance is at $t_\mathrm{p}=430$~ns (circles and dashed line as a guide for the eye). Squares and triangles were measured at $t_\mathrm{p}=180$ and 650~ns, respectively.
	\label{fig:Q-resonance}	}
\end{figure}

The scaling law described above provides a powerful tool for choosing optimal parameters for experiments using QR. To illustrate this we carry out experiments to observe resonant momentum transfer to atoms as $T$ is scanned across $T_T$. For chosen values of $N=10$ and $V_{\mathrm{d}}/h=5.89$~MHz, the scaling law predicts that the largest $\Delta P_{\mathrm{max}}$ on resonance is achieved for $t_\mathrm{p} \sim 430$~ns. Fig.~\ref{fig:Q-resonance} shows measured data with these parameters. For comparison, data at other values of $t_\mathrm{p}$ are also plotted (180 and 650~ns). We see that the largest momentum transfer as well as highest (relative to its baseline) and narrowest peak occur at $t_\mathrm{p} = 430$ ns, as expected. From a measured Talbot time ($T_T=64.8$~$\mu$s) one can deduce $h/M$, where $h$ is Planck's constant. High precision determination of $h/M$ is of general interest, as together with other well known constants it constitutes a measurement of the fine structure constant \cite{Cadoret2008}. 

\section{Discussion and conclusion}
The maximum momentum difference of the atomic distribution measured at resonance in Fig.~\ref{fig:Q-resonance} is $\Delta P_\mathrm{max} = 202 \hbar k_L$. If one uses QR as a BS in an atom interferometer, then $\Delta P_\mathrm{max}$ measures the momentum difference between the interferometer arms.  
In comparison, the state-of-the-art schemes of LMT BS are typically reaching lower values \cite{Mueller2008,Kasevich2011,Mueller2009}. It has to be noted, that the measurement in Fig.~\ref{fig:Q-resonance} can be interpreted as an atom interferometer itself since QR is a matter-wave interference effect. 
We would like to point out that we use short pulses compared to Bragg diffraction schemes, which is beneficial for avoiding incoherent photon scattering events. On the other hand, we use pulse durations above the validity range of the $\delta$-kicked particle approximation. Interestingly \cite{Altin2013} found that when operated in the quasi-Bragg regime (using $t_\mathrm{p}$ too short to fulfill the Bragg condition), their Bragg based atom interferometer reached highest contrast for $T$ that gives rise to QR.  

To conclude,  we have shown that with just 10 pulses we can generate momentum differences of around 200 $\hbar k_L$. Using QR with finite duration pulses is therefore a promising scheme for a LMT BS that may be applicable in high precision metrology. 
Furthermore, we have experimentally verified an $\epsilon$-pseudoclassical model that includes finite pulse duration for the atom optics kicked particle at QR. This model captures the quantum behavior with an effective classical treatment. 
We have found a practically useful scaling law to predict the momentum separation generated as a function of experimental parameters. Combined with the $\epsilon$-pseudoclassical model this is a powerful tool to choose optimum parameters for atom interferometry based on QR.

\section*{ACKNOWLEDGEMENTS}
We acknowledge support from the NZ-MBIE (contract No.~UOOX1402), the Leverhulme Trust (Grant No.~RP2013-K-009), and the Royal Society (Grant No.~IE110202). We thank I.~G.~Hughes for useful discussions.

\section*{APPENDIX A: INITIAL ATOMIC DISTRIBUTIONS} 
\noindent The initial momentum distribution was determined from the time-of-flight measurements. 
The initial position distribution was estimated from reverse extrapolation of time-of-flight measurements. The width of this distribution varied up to $20\%$ over the measurements.

\section*{APPENDIX B: DIPOLE POTENTIAL DEPTH}
\subsection{Calculation of the dipole potential depth}
\noindent The potential depth $V_\mathrm{d}$ is determined from the light shift (AC Stark shift) on the ground state of the atoms caused by the $\sim3$~GHz red-detuned linearly polarized standing wave (SW) light beam. In our experiments the $^{85}$Rb atoms are prepared in the F$=2$ ground state and the SW light is 40~MHz red-detuned from the F$=3$ to F'$=4$ transition. Due to the close vicinity of the D$_2$ line we only include transitions on this line in the calculation of the light shift. Using the dipole matrix elements $\mu_{m_\mathrm{F}j}$ between the ground state (F$=2$) and the multiple excited states (F'$=1,2,3$) and the detuning values $\Delta_{m_\mathrm{F}j}$, the dipole potential can be expressed as follows \cite{Grimm2000}:
\begin{equation} 
U_\mathrm{m_\mathrm{F}} = \frac{I_0}{2\varepsilon_0c\hbar} \sum_j \mu_{m_\mathrm{F}j}^2 / \Delta_{m_\mathrm{F}j} .
\end{equation} 
$I_0$ is the light intensity, $\varepsilon_0$ is the vacuum permittivity, $c$ is the speed of light in vacuum. Since we are using linearly polarized light and relatively large detuning, the variation of $U$ with $m_\mathrm{F}$ is less than $1\%$ and is neglected.

To estimate the potential depth, we need to determine the light intensity in the SW beam. For this we measured the incoming beam power, beam waist, and losses on the relevant optical elements. $V_\mathrm{d}$ is the difference between the dipole potential value at the peak intensity (in the SW anti-nodes) and its value at minimum intensity (in the SW nodes). The minimum intensity is not zero due to the power mismatch between the incoming and the retro-reflected beams creating the SW beam. 

\subsection{Variation of the dipole potential depth}
\noindent We have observed several effects that may cause different atoms experience different dipole potential depths. 
The main contribution arises from the spatial variation in intensity of the SW beams. We have measured that the beams contain intensity variation of up to a factor of 2 difference between minimum and maximum values over the region that the atoms occupy. 
Furthermore, we observed SW power fluctuations of up to $10\%$ over the experimental runs. We ascribe the variation of $V_\mathrm{d}$ required for best fit to the measurement shown in Fig.~3~(c) to these imperfections.

\section*{APPENDIX C: NUMERICAL METHODS}
\noindent For the full quantum model the Floquet operator $\hat{F}$ (Eq.~(1)) is applied $N$ times (number of SW pulses) to a momentum eigenstate. 
For a thermal atomic distribution we first calculate the momentum space wave function using the Floquet operator for a range of initial momenta spanning from $-160$ to 160 photon recoil momenta. Then we average the momentum space probability densities, each weighted with the probability for the initial momentum found from a Maxwell-Boltzmann distribution with the experimentally measured temperature. This yields the momentum distribution that is used to determine the spatial distribution from an initial point source after time-of-flight. This is convolved with the initial spatial distribution of the atomic cloud to get the final atomic distribution. The momentum distribution in Fig.~3 is obtained by converting the spatial coordinate ($x$) to momentum by $p = x M / t_\mathrm{TOF}$, where $M$ is the atomic mass and $t_\mathrm{TOF}$ is the time-of-flight. 

For the $\delta$-kicked particle the $\hat{F}$ operator was simplified by the following. (i) The $\hat{p}^2/2M$ term is neglected during the interaction with the SW pulses and (ii) the free evolution term is applied for a time $LT_T$ instead of $LT_T-t_\mathrm{p}$. The distribution of the atomic cloud is calculated following the same steps as for the full quantum model.

In the $\epsilon$-pseudoclassical model \cite{Gardiner2016} we averaged the outcomes for a large number of atomic trajectories (typically $10^5$) with initial conditions randomly sampled from the initial momentum and position distributions. 
For all subfigures of Fig.~2 we used initial momentum of $\beta \hbar K$ and for the classical and the $\epsilon$-pseudoclassical models a flat distribution of position over the spatial period of the SW.

\section*{APPENDIX D: DATA ANALYSIS}
\noindent To determine the maximal momentum width we applied smoothing to the measured momentum distribution in order to suppress noise fluctuations in the regions without atoms. This was done using a moving average filter with a span of $4\hbar k_{L}$, using Matlab's default smoothing function 5 times.

\section*{APPENDIX E: THE SCALING LAW FOR A FINITE TEMPERATURE GAS}
\noindent 
Our aim is to find and experimentally verify a scaling law that helps to optimize the experimental parameters for large momentum transfer. 
The free parameters are $N, V_{\mathrm{d}}$, and $t_\mathrm{p}$. 
$N$ and $V_{\mathrm{d}}$ are typically constrained, their ideal choice is thus straightforward (for example for $V_{\mathrm{d}}$ the optimal choice is to use the maximal laser power available). The optimal value for $t_\mathrm{p}$ is non-trivial. We therefore wish to use the scaling law to determine it. 
To do so, we need to modify the scaling law, i.e.\ the $\langle\mathcal{J}^2/2\rangle$ versus $N\tilde{V}$ function (shown as the inset of Fig.~5), such that the vertical axis contains $\Delta P_{\mathrm{max}}$ but is independent of $t_\mathrm{p}$ (see definitions of $\mathcal{J}, N, \epsilon$ and $\tilde{V}$ in the main text.) 
Since we use a thermal gas, a measurement of the mean kinetic energy or $\langle\mathcal{J}^2/2\rangle$ would be skewed by the large proportion of atoms with $\beta$ away from resonances. For a wide range of parameters $\Delta P_{\mathrm{max}}$ (which was introduced to be the maximal width of the momentum distribution) is dominated by the resonant atoms, so it is a good choice to search for an equivalent scaling law expressed in terms of $\Delta P_{\mathrm{max}}$. 
We assume that $\Delta \mathcal{J}_{\mathrm{max}} \equiv \Delta P_{\mathrm{max}} K t_\mathrm{p}/M$ is proportional to $\sqrt{\langle\mathcal{J}^2/2\rangle}$ (for the $\beta =0$ subspace).  This is naturally also a universal function of $N\tilde{V}$, but $\Delta \mathcal{J}_{\mathrm{max}}$ contains $t_\mathrm{p}$. 
The vertical axis of the universal curve can be multiplied or divided by any function of the horizontal scale, while still remaining a universal curve. To find a scaling law with the vertical axis independent of $t_\mathrm{p}$, we divide $\Delta \mathcal{J}_{\mathrm{max}}$ by $\sqrt{N \tilde{V}}$ $ = K   t_\mathrm{p} \sqrt{N V_{\mathrm{d}} /2M}$. In this way the value of $t_\mathrm{p}$ that maximizes $\Delta P_{\mathrm{max}}$ for any given $V_{\mathrm{d}}$ and $N$ can be determined from the peak of the graph. 

Fig.~5 shows the experimentally motivated scaling law. The vertical axis is proportional to $\Delta P_{\mathrm{max}}$, and it is given in units of square root of time. For our choice of units we have omitted a constant $\sqrt{2/M}$ and divided the expression for $\Delta \mathcal{J}_{\mathrm{max}}$ by $P_\mathrm{r}/\sqrt{h}$, such that $\Delta P_{\mathrm{max}}$ is in units of $ P_{\mathrm{r}}$ and  $V_{\mathrm{d}}$ is in units of frequency. This expression is independent of $t_\mathrm{p}$. Thus, we can use Fig.~5 to determine the optimal value of $t_\mathrm{p}$ for given $V_{\mathrm{d}}$ and $N$.


\begin{thebibliography}{00}
\bibitem{J} Note, that $\hat{\theta}$ and $\hat{\mathcal{J}}$ are rescaled position and momentum operators, rather than angle and angular momentum operators.
\end{thebibliography}
\end{document}